\documentclass[letterpaper]{article} 
\pdfoutput=1
\usepackage[preprint]{aaai2027}  
\usepackage[hyphens]{url}  
\usepackage{graphicx} 
\usepackage{natbib}  
\usepackage{caption} 
\usepackage{algorithm}
\usepackage{algorithmic}

\usepackage{newfloat}
\usepackage{listings}
\DeclareCaptionStyle{ruled}{labelfont=normalfont,labelsep=colon,strut=off} 
\floatstyle{ruled}
\newfloat{listing}{tb}{lst}{}
\floatname{listing}{Listing}

\usepackage{booktabs}

\usepackage{amsmath}
\usepackage{multirow}

\title{Competitive Market Behavior of LLMs}
\author{
    Pawel Struski\textsuperscript{\rm 1,\rm 3,\rm 4}\corresponding,
    Jakub Swistak\textsuperscript{\rm 2},
    Inez Okulska\textsuperscript{\rm 3},
    Przemyslaw Biecek\textsuperscript{\rm 1,\rm 2,\rm 3}
}
\affiliations{
    \textsuperscript{\rm 1}University of Warsaw\\
    \textsuperscript{\rm 2}Warsaw University of Technology\\
    \textsuperscript{\rm 3}Centre for Credible Artificial Intelligence (CCAI)\\
    \textsuperscript{\rm 4}Group for Research in Applied Economics (GRAPE)\\
    p.struski@uw.edu.pl
}

\lstdefinestyle{promptcol}{
  basicstyle=\footnotesize\ttfamily,
  breaklines=true,
  breakatwhitespace=true,
  columns=fullflexible,
  keepspaces=true,
  showstringspaces=false,
  numbers=none,
  xleftmargin=0pt,
  framexleftmargin=0pt
}

\begin{document}

\maketitle

\begin{abstract}
Large language models (LLMs) are increasingly deployed as economic agents, yet there is little evidence whether LLM agents are suited for participating in market mechanisms designed for humans, and whether these mechanisms deliver desired outcomes when faced with LLM agents. We address this question by replicating seminal economic experiments, replacing human subjects with LLM agents. We place agents in a double auction environment, which is a widely-used market mechanism. We check whether such a market is able to deliver an efficient allocation of resources, thereby testing a novel dimension of alignment of LLM agents -- their compatibility with a fundamental market mechanism. We find that markets populated by LLM agents exhibit slower or no convergence towards market equilibrium, thus providing less efficient allocations than markets populated by humans. We then analyze agents' individual trading decisions and find substantial heterogeneity both across model families and market roles. We also run a lexical analysis of Chain-of-Thought (CoT) traces generated by the agents. We find that the decision to execute a trade rather than continue incrementally adjusting prices is associated with a shift from strategic considerations toward urgency. We publicly release our testing framework\footnote{\url{https://github.com/jswistak/competitive-market-simulation}}, which can be used for future evaluations.

\end{abstract}


\section{Introduction}

Large Language Models (LLMs) increasingly act as autonomous agents that accomplish a wide range of tasks \citep{wang2024survey}, and a growing share of those tasks are economic: LLMs negotiate \citep{bianchi2024llmnegotiation}, price products \cite{fishAlgorithmicCollusionLarge2025}, bid for items \citep{chenPutYourMoney2024}, and conduct financial tasks \citep{yuFinConSynthesizedLLM2024}. In parallel, LLMs are increasingly used to simulate human behavior \citep{parkGenerativeAgentsInteractive2023, Argyle_Busby_Fulda_Gubler_Rytting_Wingate_2023}. This raises the question to what extent the behavior of LLM agents resembles human behavior. Existing research has so far focused on individual behavior \cite{hortonLargeLanguageModels2026} or relatively simple interactions in stylized-scenarios \cite{akataPlayingRepeatedGames2025}. In contrast, there has been relatively less research devoted to understanding whether agents resemble humans when faced with specific market mechanisms, which aggregate simple interactions between agents into more complex collective outcomes. We study this question in a \emph{double auction}, which is a foundational market institution with sharp theoretical predictions \citep{friedman2018double}. We ask whether LLM agents trading autonomously in a double auction converge to the competitive market equilibrium, as human traders famously do.


\paragraph{The Significance of the Double Auction.} The double auction is used in many real-world markets. The vast majority of centralized financial markets around the world are organized as a double auction. Examples include: major stock exchanges (e.g. the NYSE); commodity-linked derivatives exchanges (e.g. Chicago Mercantile Exchange); wholesale electricity markets (e.g. European Power Exchange).
The double auction is so widely used because of its allocative efficiency. In equilibrium, goods are allocated to the buyers who value them most from the sellers who can supply most cheaply. This happens even when no participant knows the equilibrium price. The double auction thus aggregates dispersed private information into a trustworthy public price without any central auctioneer, thereby solving a coordination problem.

Therefore, understanding whether the double auction can deliver this efficient outcome when populated by LLM agents is of crucial importance when it comes to market design. Should the answer be found to be negative, then markets populated by LLM agents might have to be organized differently than markets designed to be used by humans.

\paragraph{Market-Institution Alignment.} In light of the above, our research question can also be viewed as a question about LLM alignment, though the dimension of this alignment is different than the ones usually discussed in the literature \cite{ji2025alignmentsurvey}. This alignment is about whether LLM agents are capable of achieving the outcomes that a certain market institution can deliver for humans. If the answer to that question is positive, then we would consider the LLM to be aligned with that market institution. Otherwise, we would say that the LLM is not aligned. Defined as such, our study checks whether some of the modern LLMs are aligned with the double auction.

\paragraph{Behavioral Experiments.} The central question is whether a market actually reaches the equilibrium. If it does, the available surplus is fully realized; if not, some mutually beneficial trades go unconsummated and gains are left on the table. Crucially, the efficiency argument offers no guarantee of convergence, as it presumes equilibrium instead of explaining how a market of self-interested, privately-informed traders would reach it without any central coordination. \citet{smithExperimentalStudyCompetitive1962} designed a series of now-famous laboratory experiments to show how markets of human traders, each knowing only their own valuation or cost, can indeed autonomously converge to the competitive equilibrium and thereby realize the maximum attainable surplus.

However, there is no a priori reason LLM agents should reproduce this decentralized result. To answer this question, we replicate the experiments of \citet{smithExperimentalStudyCompetitive1962}, replacing human subjects with LLM agents across different model families and capability tiers. We analyze both the emergent market-level dynamics and the individual behavior of our agents. Our framework studies actual decision-making rather than stated preferences yet remains highly tractable.

\paragraph{A Clean Experimental Environment.} A growing literature has proposed various benchmarks for LLM agents to test how they perform in economic tasks (\citealp{yuFinConSynthesizedLLM2024}; \citealp{xieFinBenHolisticFinancial2024}; \citealp{ramanSTEERAssessingEconomic2024b}; \citealp{backlundVendingBenchBenchmarkLongTerm2025a}; \citealp{fishEconEvalsBenchmarksLitmus2025}). While undoubtedly useful, these benchmarks can be difficult to interpret and often rely on highly specific design decisions that may not generalize across environments. More broadly, beyond economic settings, multi-agent LLM studies that examine emergent collective behavior often rest on qualitative observations rather than on quantifiable, falsifiable metrics \citep{la2026large}. \textbf{Our setting avoids both limitations.} We draw on a canonical economic experiment, backed by decades of empirical research. In our setting, equilibrium predictions, allocative efficiency, and convergence are all precisely measurable and benchmarked against both theory and human experimental data.


Our contributions can be summarized in the following way:
\begin{enumerate}
  \item We introduce a re-usable framework for simulating Smith-style double auction experiments with LLM agents, providing a tractable environment with clear human and theoretical benchmarks, in which outcomes are precisely measurable.
  
  \item We show that convergence to the competitive market equilibrium is not guaranteed, and that markets with LLM agents exhibit worse convergence than markets populated by humans, which suggests LLM agents may exhibit a certain degree of misalignment with the market institution of the double auction.
  
  \item We provide a quantitative behavioral characterization of LLM traders, documenting substantial heterogeneity both across model families and across market roles. For a selected experiment, we supplement this with a linguistic analysis of the models' internal reasoning tokens.
  
\end{enumerate}

\section{Related Work}

\paragraph{LLM Agents in Markets.} A growing body of work studies the economic behavior of LLM agents in
markets, from individual decision-making to the market-level phenomena that emerge when many agents interact. Among these, studies that replicate controlled laboratory experiments concentrate largely on bubbles
and asset markets \citep{saxena2026machinespiritsspeculationadaptation,
rio-chanonaCanGenerativeAI2025, henningLLMAgentsNot2025}, whereas those targeting more general market mechanisms and competition either examine
narrow, specific markets without an explicit human benchmark
\citep{zhaoCompeteAIUnderstandingCompetition2024a}, or restrict attention to
agents' bidding strategies \citep{chenPutYourMoney2024}, leaving market efficiency unaddressed. Closer to our setting,
\citet{shahLearningSyntheticLabs2025} study several common auction formats but omit the double auction, and therefore observe only the buy side of the market. The active role of both buyer and seller in the double auction lets us compare the
buy and sell sides directly, a distinction that turns out to be substantive. We adopt the deliberately simplified laboratory setting of
\citet{smithExperimentalStudyCompetitive1962} to isolate a fundamental market mechanism in a clean, general form. Our study brings together three elements
that, to our knowledge, no prior work combines: a controlled environment grounded in experimental economics with a human benchmark; an explicit treatment of market efficiency rather than agent performance alone; and a two-sided design that captures both the buyer and seller side.

\paragraph{Convergence to Competitive Equilibrium.} A replication of the market convergence experiments of
\citet{smithExperimentalStudyCompetitive1962} has already been conducted by \citet{jiaExperimentalStudyCompetitive2024}, who report that LLM agents do not converge to the competitive equilibrium. We revisit this question and reach a more nuanced conclusion: while we have not observed complete convergence in any of our experiments, we observe evidence of partial convergence in some of them (see Section~\ref{sec:market-wide}). Crucially, our study differs from theirs in a few significant ways. First, we evaluate different models and average over repeated runs to account for the stochasticity of LLM behavior. Second, our implementation of the double auction mechanism is different. Whereas they prompt agents to post prices that are simply recorded and matched whenever a bid and ask cross, we implement a persistent order book, in which only spread-improving quotes are admitted and a crossing quote executes immediately against the resting order (see Section \ref{sec:trading_mechanism}). Finally, beyond the convergence result, we provide a more detailed analysis of agent behavior, explicitly quantify market efficiency, and release our testing framework, which supports evaluating different models across a range of market conditions.

\paragraph{Market Collusion of LLM Agents.} Our work also connects to a line of work that studies a specific form of misalignment with market institutions, namely, the collusion among LLM-based pricing agents \citep{fishAlgorithmicCollusionLarge2025, linStrategicCollusionLLM2025, agrawalEvaluatingLLMAgent2025}. Such studies often interpret supracompetitive prices and incremental undercutting as evidence of collusion. Our results bear directly on this reading: because we find that LLM traders do not reliably converge to competitive equilibrium, such signals are harder to attribute cleanly to collusion than standard economic reasoning might suggest. See Section~\ref{sec:discussion}. 

More broadly, with the exception of the market collusion literature, multi-agent alignment has received comparatively little attention \citep{carichon2025comingcrisismultiagentmisalignment}. Viewed through the alignment lens we adopt, collusion is one particular failure of market-institution alignment. Our study steps back from this single mechanism to the more general question of whether that alignment is obtained at all i.e. whether markets populated by LLM agents attain their intended allocative efficiency, to which collusion is only one of several possible obstacles.

\paragraph{Agent-Based Models.} Our work has a bearing on the literature about Agent-Based Models (ABMs) in economics \cite{tesfatsion2006handbook}, but our approach is fundamentally different than that of traditional ABMs. Our goal is to characterize LLM agents themselves (their behavior, and their alignment with the double auction mechanism) using the experimental methods of behavioral economics. ABMs run the other way. ABMs take agents as given, typically with fixed decision rules, and simulate their interactions to learn about the economy. Thus, in ABMs, agents are instruments while the economy is the object of study. For us, the agents' behavior is the object of study. That said, to the extent that LLMs are found to be good approximations for human behavior, which remains an open question, ABMs that rely on LLM agents instead of rules-based agents could prove to be useful \citep{gaoLargeLanguageModels2023, liEconAgentLargeLanguage2024a}. Our work sheds light on this topic by characterizing the behavior of LLM agents relative to humans in the fundamental market institution of double auction.

  \section{Method}

  We develop a framework for simulating continuous double auction (CDA) markets
  populated by LLM agents.

\subsection{Market Environment}
\begin{figure}[t]
    \includegraphics[width=\columnwidth]{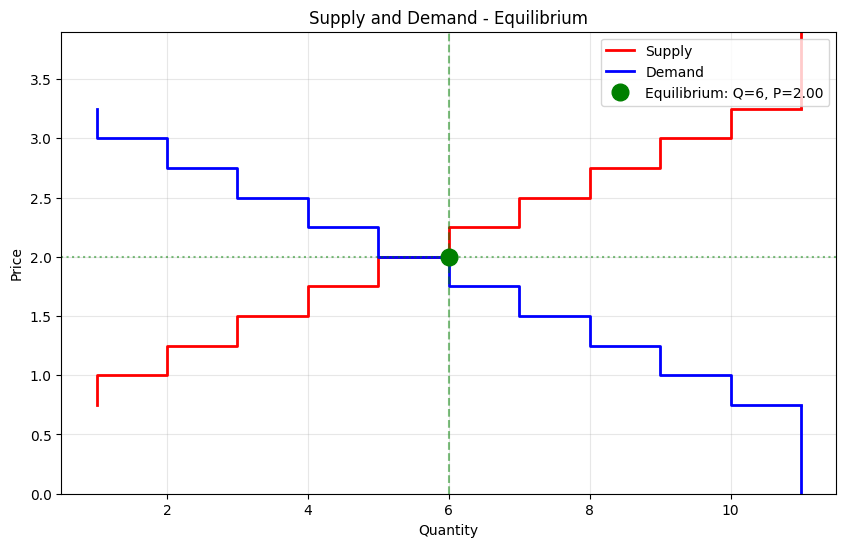}
    \caption{\textbf{Market design.} Reservation prices induce symmetric step-function demand and supply schedules. The competitive equilibrium $(p^\star, q^\star) = (\$2.00,\, 6)$ lies at their intersection.}
    \label{fig:market-equilibrium}
\end{figure}

The market contains two types of agents, buyers and sellers, trading a single unspecified
good. Each agent $i$ is endowed with a private reservation price $v_{i}$: for buyers,
the maximum price at which they can purchase; for sellers, the minimum price
at which they can sell\footnote{Although reservation prices bound the range of profitable transactions, the trading mechanism imposes no such constraint: agents are
free to post and accept offers that yield negative profits.}. We use 11 buyers
and 11 sellers with reservation prices ranging from \$0.75 to \$3.25 in \$0.25
increments, producing symmetric, step-function demand and supply schedules
that intersect at a competitive equilibrium price $p^\star = \$2.00$ and equilibrium
quantity $q^\star = 6$ trades per round (Figure \ref{fig:market-equilibrium}). 

This intersection defines the \emph{competitive equilibrium}: the price at which
the quantity buyers are willing to purchase equals the quantity sellers are willing
to supply, and at which total surplus -- the sum of realized buyer and seller
rents -- is maximized. Crucially, $p^\star$ is a prediction derived from the
full demand and supply schedules, yet no individual agent observes these
schedules or is told that \$2.00 is the equilibrium price; each agent knows
only its own reservation value. The central finding of decades of experimental economic research shows that decentralized trading among such privately-informed,
self-interested agents drives transaction prices toward this competitive prediction over repeated rounds \citep{davis1993experimental}. A market-level equilibrium thus emerges from purely local behavior.

Our market follows the symmetric baseline design of \citet{smithExperimentalStudyCompetitive1962}. We adopt a symmetric specification deliberately. Since the supply and demand schedules are mirror images, any asymmetry we observe between buyers and sellers can be attributed to model-specific behavior rather than to the market design.

\subsection{Trading Mechanism}
\label{sec:trading_mechanism}
Each experiment consists of 5 rounds. In each round, every agent is endowed with
one unit of supply (sellers) or one unit of demand (buyers) and may execute at
most one transaction. The market maintains a standing best bid $b^\star_t$ (the highest outstanding
buy offer) and a standing best ask $a^\star_t$ (the lowest outstanding sell offer); either
may be absent if no order has been posted on that side.

A round proceeds in up to $T = 300$ iterations. At each iteration $t$, an agent $i$ is selected
uniformly at random from those who have not yet transacted in the current round,
and is given the opportunity to post an offer: a bid $b_{i,t}$ if $i$ is a buyer or an ask $a_{i,t}$
if $i$ is a seller. Offers are processed as follows (stated for a buyer; the seller case is symmetric):
\begin{itemize}
    \item \textbf{Crossing the spread.} If $b_{i,t} \geq a^\star_t$, a transaction executes immediately
        at price $p_t = a^\star_t$, and both parties exit the round.
    \item \textbf{Improving quote.} If $b^\star_t < b_{i,t} < a^\star_t$, the bid becomes the new
        standing bid.
    \item \textbf{Rejection.} If $b_{i,t} \leq b^\star_t$, the bid is rejected by the exchange
        and has no effect.
\end{itemize}
A round terminates when the iteration cap $T$ is reached.

\subsection{Information and Objective}
At each decision point, the selected agent $i$ acts on the information set
\begin{equation}
    \mathcal{I}_{i,t} = \bigl\{\, v_i,\; b^\star_t,\; a^\star_t,\; \mathcal{H}_t,\; \mathcal{H}_{i,t} \,\bigr\},
\end{equation}
comprising its own reservation price $v_i$, the current standing bid $b^\star_t$ and ask $a^\star_t$, the market-wide history $\mathcal{H}_t$ of all posted, rejected, and accepted offers in the current experiment through iteration $t$, and its own private history $\mathcal{H}_{i,t}$ of past actions and their outcomes. After each iteration, a new row is appended to $\mathcal{H}_t$ and, if the selected agent acted, to $\mathcal{H}_{i,t}$.

Agents are instructed to maximize profit. Denoting by $p_t$ the transaction price
at which agent $i$ trades, realized profit is
\begin{equation}
    \pi_i =
    \begin{cases}
        v_i - p_t & \text{if $i$ is a buyer}, \\
        p_t - v_i & \text{if $i$ is a seller},
    \end{cases}
\end{equation} 
and $\pi_i = 0$ if $i$ does not transact. No instruction is given
regarding the trade-off between transaction probability and per-trade profit, nor are agents informed of the competitive equilibrium price\footnote{Detailed prompts are shown in the supplementary material.}.

\subsection{Experimental Conditions}

We conduct the above market experiments with LLM agents based on the current high-capability models offered by major providers. In our experiments, we have gpt-5.4-2026-03-05,  gpt-5.4-mini-2026-03-17 \cite{openai_api_2026}, and gemini-3.1-pro-preview \cite{gemini_api_2026}, which we access via their respective APIs. We dub them GPT Large, GPT Small, and Gemini Large respectively. Each experiment has been ran 10 times independently with 10 different random seeds to account for the
stochastic nature of our experimental design as well as the stochasticity of
LLMs. The parameters of all LLMs have been kept at the default level. Notably,
the temperature parameter has been set to 1. 

The code needed to replicate those experiments, along with the appropriate configuration files can be found on GitHub\footnote{\url{https://github.com/jswistak/competitive-market-simulation}}. A valid API key is required to access each of the tested LLMs. The data collected during our experiments is also released in the supplementary material, along with the code that was used to analyze that data. The code was run on a laptop with a CPU and 32GB of RAM.

\section{Results}
\label{sec:results}

\subsection{Market-Wide Outcomes}
\label{sec:market-wide}
Table \ref{tab:smith-convergence} reports our results together with the original human-subject benchmark from \citet{smithExperimentalStudyCompetitive1962}. It shows per-round means of the number of trades, transaction prices, price dispersion around the competitive equilibrium ($\alpha$), allocative efficiency (Eff.). All metrics are averaged across 10 simulations per condition. 

The price dispersion $\alpha$ is defined as the ratio of the standard deviation of transaction prices around the equilibrium price $\sigma_0$ to that equilibrium $P_0$ price, expressed as percentage i.e. $\alpha = 100 \cdot \frac{\sigma_0}{P_0}$. This is the main number that we compare against the human benchmark. 
 
Allocative efficiency is the total realized surplus
as a share of the maximum surplus attainable in equilibrium i.e. the sum of all buyers' and sellers' realized profits divided by the total gains from trade available when the equilibrium quantity is exchanged. It equals 1 when every profitable trade is executed and falls below 1 whenever gains from trade are left on the table, whether because profitable trades go unexecuted or because units are misallocated to lower-value buyers or higher-cost sellers.
 
\begin{table}[t]
\centering
\setlength{\tabcolsep}{4pt}
\small
\begin{tabular}{@{}p{1.3cm}ccccc@{}}
\toprule
Model & R & Trades & Price & $\alpha$ & Eff. \\
\midrule
\multirow{5}{=}{Gemini Large} & 1 & 5.4 & 1.94 & 24.7 & 0.83 \\
                              & 2 & 4.6 & 2.26 & 20.6 & 0.70 \\
                              & 3 & 3.9 & 2.18 & 25.7 & 0.66 \\
                              & 4 & 3.3 & 2.26 & 20.9 & 0.58 \\
                              & 5 & 4.1 & 2.24 & 28.1 & 0.72 \\
\midrule
\multirow{5}{=}{GPT Large}    & 1 & 2.6 & 2.00 & 20.1 & 0.47 \\
                              & 2 & 3.8 & 1.87 & 22.1 & 0.59 \\
                              & 3 & 3.0 & 1.96 & 18.6 & 0.56 \\
                              & 4 & 2.2 & 2.13 & 18.8 & 0.36 \\
                              & 5 & 3.5 & 1.82 & 15.6 & 0.67 \\
\midrule
\multirow{5}{=}{GPT Small}    & 1 & 5.5 & 2.33 & 27.6 & 0.79 \\
                              & 2 & 5.7 & 2.29 & 18.2 & 0.86 \\
                              & 3 & 5.7 & 2.20 & 12.6 & 0.89 \\
                              & 4 & 5.8 & 2.19 & 13.6 & 0.89 \\
                              & 5 & 6.2 & 2.13 & 11.7 & 0.91 \\
\midrule
\multirow{5}{=}{Smith (1962)} & 1 & 5 & 1.80 & 11.8 & -- \\
                              & 2 & 5 & 1.86 & 8.1  & -- \\
                              & 3 & 5 & 2.02 & 5.2  & -- \\
                              & 4 & 7 & 2.03 & 5.5  & -- \\
                              & 5 & 6 & 2.03 & 3.5  & -- \\
\bottomrule
\end{tabular}
\caption{Equilibrium convergence metrics across experiments. Each block corresponds to a different LLM model agent population, with five trading rounds (R) per experiment. Equilibrium quantity is 6 and equilibrium price is 2.0 in all conditions. Metrics: Trades (number of trades), Price (transaction price), $\alpha$ (price dispersion around equilibrium), Eff. (allocative efficiency). All LLM metrics are mean values across the 10 simulations.}
\label{tab:smith-convergence}
\end{table}

\paragraph{There is No Full Convergence.}
In none of the three experiments do we observe full convergence to the
competitive equilibrium of $p=2.00$, $q=6$. Across all three model
populations, the coefficient of convergence $\alpha$ remains well above the
human benchmark in every one of the five trading rounds. More tellingly, the
\emph{shape} of the series differs: in the human data $\alpha$ trends mostly steadily
downward (from 11.8 in round 1 to 3.5 in round 5), whereas in two of the three
experiments $\alpha$ oscillates from round to round with no clear or sustained
downward trend, and only one (GPT Small) reproduces the human pattern of a declining
$\alpha$. Trade volume falls short of the equilibrium quantity of 6 in every round but one (GPT Small, round~5). Allocative efficiency tells the same story: no condition reaches full
efficiency in any round, with even the best-performing condition peaking at
0.91, and the least efficient markets falling well below that. This means that the market fails to fully realize the available gains from trade.

\paragraph{The Smallest Model Converges Most.}
GPT Small is the condition in which we observe most significant convergence toward the competitive equilibrium of p=2.00, q=6. Across rounds, mean prices fall from \$2.33 to \$2.13, price dispersion ($\alpha$) declines from 27.6 to 11.7, and allocative efficiency rises from 0.79 to 0.91. Convergence is incomplete relative to Smith's human subjects — even in round 5, dispersion is roughly 3 times the human value (11.7 vs. 3.5) and the mean price remains \$0.13 above equilibrium, much higher than the \$0.03 in the human benchmark. Nonetheless, this is the condition in which the convergence pattern is most visible.

It is notable that the smaller, less capable model produces more efficient markets than any of the two larger ones. That said, our experimental design is not indented to provide evidence regarding the effects of model scale, so a proper examination would require a separate study. Nonetheless, we find this result surprising and counter-intuitive at first, and we return to it in Section \ref{sec:discussion}. 

\paragraph{Aggressive Rent-Seeking Suppresses Trade Volume.}
The number of trades in the GPT Large condition is substantially below other conditions and well below the equilibrium quantity of 6, averaging 2.2–3.8 trades per round. This reflects the reluctance of GPT Large agents to cross spreads in larger increments (a behavioral pattern we generally examine in Section~\ref{sec:individual}, and which is mostly prounced in the case of GPT Large). Agents spend rounds making \$0.01 improvements without ever closing the spread, and the round terminates before trades can occur. The result is depressed allocative efficiency: 0.36–0.67 across rounds, which is on average lower than in the case of the two other models. This is a clean illustration of how a single behavioral feature at the individual level (unwillingness to cross spreads) translates into large differences in market-wide outcomes. Admittedly, extending the iteration cap could increase the number of transactions within each round in this experiment. However, even if it were the case, the speed of convergence is itself a substantive outcome. The allocative efficiency of a double auction is fully realized only in equilibrium. A market that takes longer to get there is, for the whole of that path, less efficient and its price signal less trustworthy. The finding that some models converge more slowly than others, or more slowly than humans, is therefore a meaningful result, even if they would all converge eventually.

\subsection{Individual Behavior}
\label{sec:individual}
Before turning to model-specific behavior, we briefly describe the typical dynamics of a single experiment run. The first agent to act posts an offer trying to secure some reasonable rent, with the exact value varying widely across runs. From there, agents move in small increments undercutting or overbidding the previous quote, until the spread becomes small enough for an agent to cross it and execute a trade. This pattern repeats across iterations and rounds. We document key findings  below.

\paragraph{Incremental Order Improvement.}
The vast majority of order improvements across our experiments are in small \$0.01 increments. Agents improve standing orders by the smallest possible amount, ensuring their offer becomes the new standing quote while preserving as much potential profit as possible. This behavior carries a cost: small improvements are more likely to be undercut by other agents, and the round may finish before the agent transacts. Agents thus face a profit-maximization vs. transaction-completion tradeoff, and our results suggest they heavily prioritize the former with little regard for the risk of failing to trade. 

\paragraph{Heterogeneous Willingness to Cross the Spread.}
Models differ substantially in the bid-ask spread they tolerate before initiating a trade. Gemini Large agents cross spreads of approximately \$0.50 on average, while GPT Large agents mostly wait until the spread approaches zero, with very few exceptions. This indicates that Gemini Large agents respond more readily to profitable trade opportunities and accept smaller per-trade profits, while GPT Large agents leave no potential profit on the table even when profitable trades are available. This is a direct manifestation of the profit-vs-completion tradeoff, with the two models occupying different points on the frontier. GPT Small agents are an intermediate case, with buyers crossing spreads of around \$0.25 on average while sellers mostly waiting until the spread closes.

\begin{figure}[t]
\includegraphics[width=\columnwidth]{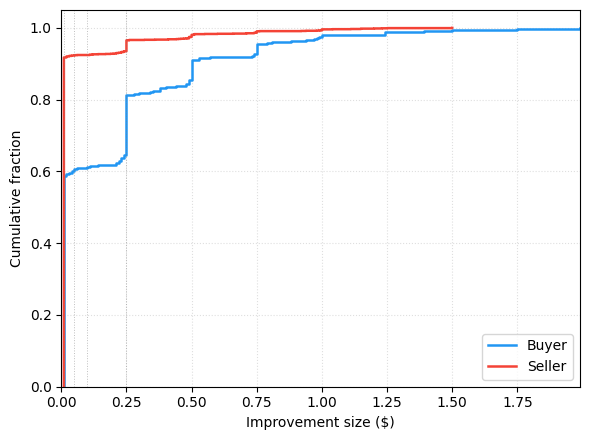
}
\caption{Empirical Cumulative Distribution Function (ECDF) of order improvement size for GPT Small agents across simulations (pooled). Small order improvements dominate among sellers, while buyers tend to place more large improvements.}
\label{fig:ecdf_gpt_small}
\end{figure}

\paragraph{Role Asymmetries.}
In GPT Small, we observe substantial behavioral differences between buyers and sellers. While over 90\% of standing-ask improvements are the incremental \$0.01 moves, nearly 40\% of standing-bid improvements exceed \$0.25 (Figure \ref{fig:ecdf_gpt_small}). This asymmetry extends to trade initiation. While the absolute number of trades initiated by each side is broadly equal, the fill rate — the fraction of an agent's orders that result in a transaction — is markedly higher for buyers (see the supplementary material). The mechanism is that sellers submit roughly three times as many orders as buyers, but many of these orders are small incremental improvements. Buyers, by contrast, often refrain from posting, and when they do post, they make larger improvements that are more likely to result in a transaction. This is a significant role-based behavioral divergence and the main exception to the incremental-improvement pattern discussed above.

\begin{figure*}[t]
\includegraphics[width=0.48\linewidth]{
  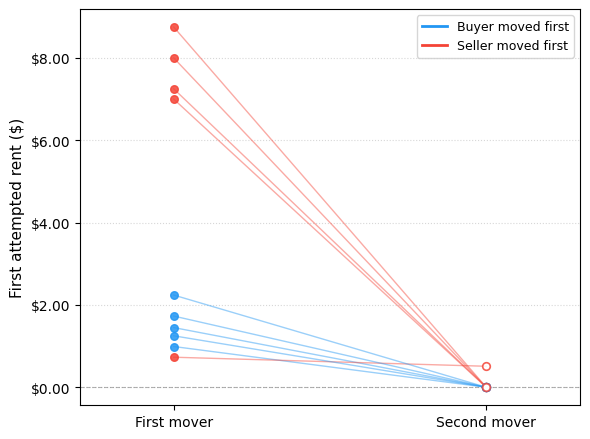
}
\hfill
\includegraphics[width=0.48\linewidth]{
  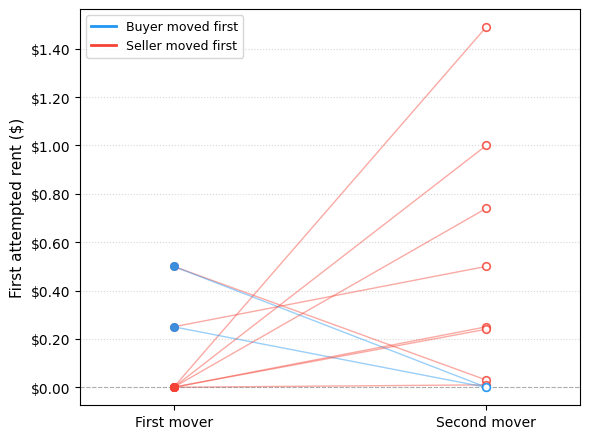
}
\caption{The first attempted rent (profit) for agents whose side is first to
place an initial order and second (individual observations). Left panel: GPT Large. In all but one case, the 2nd moving agent immediately falls to his reservation price. Right panel: GPT Small. The pattern is partly reversed.}
\label{fig:first-mover}
\end{figure*}

\paragraph{Opening-Offer Anchoring.}
The first offer placed in an experiment serves as an anchor for subsequent agents. For agents on the same side as the opening offer, this is unsurprising because, to have their own offer accepted, they must improve on the standing quote, which mechanically ties their behavior to it. What is surprising is that agents on the opposing side react to the first offer as well, even though they have yet to post any offer of their own and are under no such constraint. Nothing stops them from placing a more aggressive offer that secures a reasonable rent; if it fails to trade or is undercut by rivals, they can simply revise it in a later iteration. Yet agents often do not pursue this strategy.

This produces a systematic asymmetry between the side that moves first and the opposing side. The distribution of opening offers for the first-moving side is wide, reflecting heterogeneous strategies for extracting rent in the absence of any reference point. The distribution of opening offers for the second-moving side is much tighter and anchored on the first offer.

This effect takes an extreme form in GPT Large agents. Virtually all initial orders from the second-moving side were placed exactly at the agent's reservation price, fully conceding rent (Figure \ref{fig:first-mover}, left panel). The single exception occurred when the initial offer was immediately crossed by the opposing side and resulted in a trade. Interestingly, in GPT Small this pattern is partly reversed: in some runs the first-moving side posts at its reservation price while the second-moving side attempts to extract rent (Figure \ref{fig:first-mover}, right panel).

\section{Analysis of Reasoning of LLM Traders}
\label{sec:reasoning}
Section~\ref{sec:individual} identified willingness to cross the spread as the
behavioral lever separating functioning markets from stalled ones. We exploit the
internal thinking tokens of Gemini Large, which is the only model in our study that exposes reasoning traces, to ask what distinguishes the decision to keep
making incremental quote improvements from the decision to cross the spread and transact. Since only one model exposes its traces, we do not claim the pattern generalizes; we treat it as a window into a single mechanism we have already shown matters for efficiency, not as a claim about LLM traders in general.

\begin{figure}[t]
    \includegraphics[width=\columnwidth]{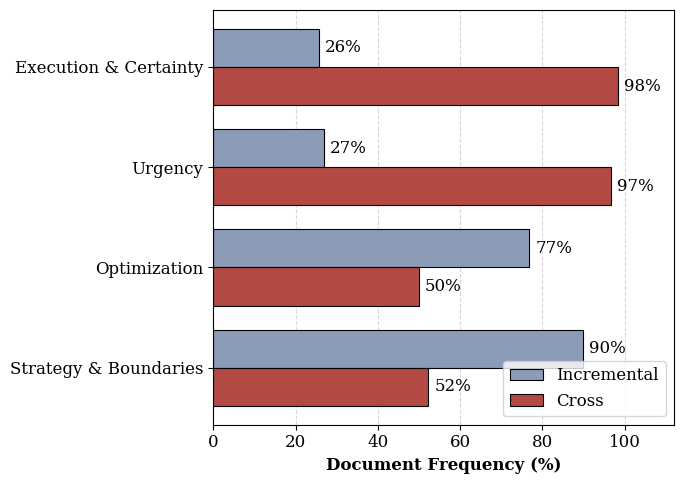}
    \caption{Lexical shift in the internal thinking between move types, based on
    terms distinctively associated with urgency, execution, optimization, or
    strategy.}
    \label{fig:cluster-shift}
\end{figure}

Gemini Large agents alternate between minimal steps (the \emph{incrementing
phase}) and larger jumps that cross the spread to secure a trade (the \emph{cross
phase}). We show how a typical experiment unfolds in the supplementary material. To characterize how the reasoning differs between the two phases, we pool the
traces from each phase and rank terms by the weighted log-odds-ratio with an
informative Dirichlet prior \citep{monroe2008fightin}, which measures how much
more characteristic a term is of crossing than of incrementing. We group the most
distinctive terms into four categories: \textbf{Urgency} (\textit{immediate,
waiting, overthink}) and \textbf{Execution \& Certainty} (\textit{secure,
guarantee, execute, lock}), associated with closing a trade; and \textbf{Strategy
\& Boundaries} (\textit{strategic, position, room, competitive}) and
\textbf{Optimization} (\textit{improve, maximize, increment, increase}),
associated with the margin-protecting reasoning of the incrementing phase. Moving
from the incrementing to the crossing phase, the first two categories rise sharply
while the latter two fall (Figure~\ref{fig:cluster-shift}).

This shift is consistent with the profit-versus-completion tradeoff of
Section~\ref{sec:individual}. The language of margin protection recedes and the
language of urgency and execution rises as agents move to trade. We read this as
agents weighting profit early and completion later. We are deliberately cautious
about a stronger interpretation. The analysis establishes which terms separate
the two phases, not what causes an agent to switch, and reasoning traces need not
faithfully reflect the computation that drives the decision. We therefore treat
these traces as suggestive rather than as direct evidence of the agents' decision process.

\section{Conclusions and Future Work}
\label{sec:discussion}

We used the experimental paradigm of \citet{smithExperimentalStudyCompetitive1962} to ask a single question: can the double auction deliver the same allocative efficiency with LLM traders as it reliably does with humans? Across three model populations spanning two providers and two capability tiers, the answer is negative. No condition converges fully to the competitive equilibrium, and price dispersion remains well above the human benchmark in every trading round. The convergence we do observe is partial and uneven: only GPT Small reproduces the human pattern of a steadily declining price dispersion. Viewed through the lens of market-institution alignment, this means that the models we study are, at best, only partially aligned with the double auction. They do not autonomously realize the efficient allocation the institution is designed to produce for humans.

Our most surprising finding is that the smallest and least capable model, GPT Small, produces the most human-like and most efficient markets, converging more closely toward the competitive equilibrium than any one of the two larger models. However, since our design was not constructed to isolate the effect of model scale, we are cautious about over-interpreting this result. A proper treatment would require an experiment built for that purpose. We nonetheless regard it as one of the more intriguing outcomes of this study, particularly as it echoes \citet{wang2025large}, who also find that larger models trade less like humans. Understanding whether this pattern actually holds under proper scrutiny, whether it emerges across other market institutions, and what might cause it, is a promising direction for future work.

Our results also carry a methodological caution for using LLMs as proxies of human behavior in ABMs. Our findings indicate that this substitution cannot always be assumed to be innocuous. Even in a setting where human behavior is well-documented and robust, LLM agents do not reproduce it off the shelf. While we do not claim that it is not possible to reproduce human behavior with LLMs, we would argue that simulations that rely on LLM agents as human stand-ins should be validated against human and theoretical benchmarks whenever these exist, rather than presumed to inherit human behavioral regularities.

Another implication of our study concerns the detection of market collusion. A recent line of work reads supracompetitive prices, together with incremental undercutting such as repeated \$0.01 ask reductions during market stasis, as evidence that LLM pricing agents collude \citep{agrawalEvaluatingLLMAgent2025}. Our results suggest caution before importing this inference from studies of human traders. Treating prices above the competitive level as collusive presumes that
non-colluding agents would converge to the competitive equilibrium in the first
place, yet in our experiments convergence is not guaranteed. To the extent that the non-cooperative benchmark for LLM traders itself sits above equilibrium, deviations above the competitive price cannot be attributed cleanly to collusion. The incremental undercutting is similarly ambiguous. It is hard to distinguish it from the anchoring on peers' visible quotes that we document in Section~\ref{sec:individual}. Establishing whether LLM markets converge in the absence of collusion is thus a prerequisite for interpreting these signals.

Finally, our results are an early step rather than a verdict on whether markets of LLM agents can function. It remains an open question whether the incomplete convergence we observe reflects a fundamental limitation or merely a slower path to equilibrium. It is conceivable that longer simulation horizons, richer feedback, or targeted interventions might lead to better convergence. To enable this investigation, we publicly release our framework for simulating Smith-style double auction experiments with LLM agents. Our framework supports evaluating additional models across a range of market conditions, and testing convergence under different supply and demand schedule would be an interesting question, which we leave for future research. Beyond the scale question raised above, natural next steps include extending the environment to other market institutions, varying the trading horizon and the information agents hold about it, and comparing a wider range of models. As LLM agents are increasingly responsible for economic decisions in real markets, establishing the conditions under which those markets remain efficient, or fail to, is an important question, which we hope our testing framework can help answer.

\section*{Acknowledgments}
We gratefully acknowledge valuable comments from Joanna Tyrowicz. This work was supported by the Polish National Science Centre under project no. 2019/34/E/ST6/00052.

\bibliography{references_clean}

\clearpage

\appendix

\section{Role Asymmetries in the Fill Rate}

\begin{figure}[t]
    \includegraphics[width=\columnwidth]{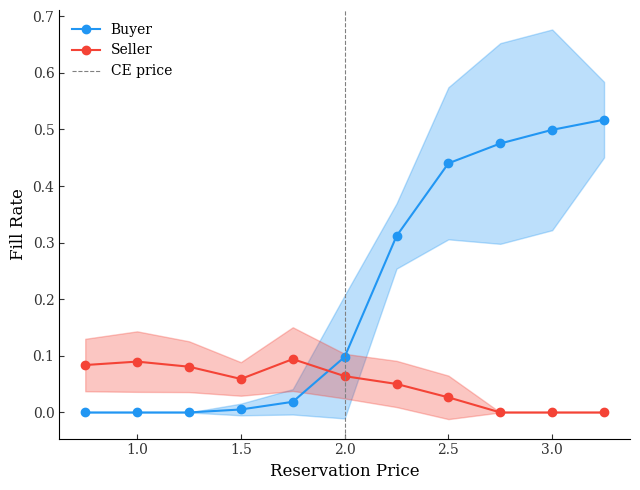}
    \caption{Fill rate (the fraction of orders that led to transactions) vs reservation price for buyer and seller agents (GPT Small). The dots indicate mean values across simulations. The shaded areas indicate 95\% confidence intervals.}
    \label{fig:fill_rate_gpt_small}
  \end{figure}

Section 4.2 of the main text mentions that we observe substantial behavioral differences between buyers and sellers among the GPT Small agents in terms of the size of order improvements they make. While agents make incremental order improvements most of the time, buyer agents tend to make larger order improvements more frequently than seller agents. This asymmetry extends to trade initiation. While the absolute number of trades initiated by each side is broadly equal, the fill rate — the fraction of an agent's orders that result in a transaction — is markedly higher for buyers (Figure \ref{fig:fill_rate_gpt_small}). The mechanism is that sellers submit roughly three times as many orders as buyers: many small incremental improvements unwilling to concede profit. Buyers, by contrast, often refrain from posting at all, and when they do post they make larger improvements that are more likely to clear. This is a significant role-based behavioral divergence and the main exception to the incremental-improvement pattern discussed above.

\section{Prompt Templates}
\label{app:prompts}

Each agent receives a static system prompt and, at every iteration, a
user prompt summarizing the market state and its own history. Both are
generated from shared templates with placeholders in braces (e.g.\
\texttt{\{role\}}) filled at runtime. Buyer- and seller-specific wording
is substituted from keywords shown in Table~\ref{tab:keywords}. Other placeholders are filled by the simulation state or config parameters.

\begin{table}[t]
\centering
\scriptsize
\setlength{\tabcolsep}{4pt}
\begin{tabular}{lll}
\toprule
Prompt Variable & Buyer & Seller \\
\midrule
\texttt{role}        & buyer  & seller \\
\texttt{verb}        & buy    & sell \\
\texttt{preference}  & lowest & highest \\
\texttt{condition}   & above  & below \\
\texttt{announcement\_type} & bid & ask \\
\bottomrule
\end{tabular}
\caption{Role-specific keyword substitutions. The
\{profit\_formula\}field is "your reservation price and your
transaction price" for buyers and the reverse order for sellers, so the
difference is signed correctly in each role.}
\label{tab:keywords}
\end{table}

\subsection{System Prompt}

\begin{lstlisting}[style=promptcol]
You are a {role} participating in a market
for a good you need to {verb}. Your task is
to {verb} a unit of the good at the
{preference} possible price but not
{condition} your reservation price.
Your profit from a transaction equals the
difference between {profit_formula}.
Your goal is to maximize your profit.
Your reservation price is known to you and
only you.

There are {N_BUYERS} buyers and {N_SELLERS}
sellers in this market (including you).

Market rules:
- The market maintains a standing best bid
(the highest outstanding buy offer) and a
standing best ask (the lowest outstanding
sell offer). Either may be absent if no one
has posted on that side yet.
- {order_outcomes}
- A round ends once no one can post an
improving order, or once a cap on orders is
reached.

There will be {N_ROUNDS} rounds. Each round,
you want to {verb} an additional unit of the
good and are able to transact irrespective
of whether you transacted in the previous
round. You can only make one transaction per
round.
\end{lstlisting}

\subsection{User Prompt}

\begin{lstlisting}[style=promptcol]
Market history:
{market_history}

History of your actions:
{own_history}

This is round {round}/{N_ROUNDS}.
Current standing bid: {standing_bid}
Current standing ask: {standing_ask}

Your reservation price is
${reservation_price:.2f}. Under no condition
can you {verb} {condition} your reservation
price.

{action_prompt}
\end{lstlisting}

\subsection{Announcement Prompts}

\begin{lstlisting}[style=promptcol]
[Buyer]
Do you want to announce a bid to buy? If so,
what is your bid price? It must be strictly
higher than the current standing bid (or be
the first bid if none exists) and no higher
than your reservation price.

[Seller]
Do you want to announce an offer to sell? If
so, what is your asking price? It must be
strictly lower than the current standing ask
(or be the first ask if none exists) and no
lower than your reservation price.
\end{lstlisting}

\subsection{Order Outcomes Block}

The \{order\_outcomes\} variable in the system prompt is
filled per role:

\begin{lstlisting}[style=promptcol]
[Buyer]
If your bid is greater than or equal to the
standing ask, a trade executes immediately
at the standing ask price. Both parties exit
the round.
Otherwise, if your bid is strictly greater
than the standing bid, your bid becomes the
new standing bid and you may transact at your
bid price if a seller later posts an ask at
or below it.
Otherwise, your bid has no effect.

[Seller]
If your ask is less than or equal to the
standing bid, a trade executes immediately
at the standing bid price. Both parties exit
the round.
Otherwise, if your ask is strictly less than
the standing ask, your ask becomes the new
standing ask and you may transact at your ask
price if a buyer later posts a bid at or
above it.
Otherwise, your ask has no effect.
\end{lstlisting}

\subsection{History Line Templates}

\begin{lstlisting}[style=promptcol]
[Market history -- accepted]
In round {round} at iteration {iteration}, an
announcement to {announcement_type} for
${price:.2f} was accepted.

[Market history -- posted, no crossing]
In round {round} at iteration {iteration}, an
announcement to {announcement_type} for
${price:.2f} was posted as the new best
{announcement_type} but no one crossed it
yet.

[Own action -- general]
In round {round}, your offer to
{announcement_type} for ${price:.2f} was
{outcome}.

[Own action -- non-improving]
In round {round}, your offer to
{announcement_type} for ${price:.2f} was
rejected because it did not improve the
standing book.
\end{lstlisting}

\section{Experimental Framework}
Our experiments were conducted in a multi-agent framework built in Python. The agent orchestration was done using LangGraph. Running the simulations requires access to each of the model provider's API (Gemini, OpenAI). The code was run on a laptop with a CPU and 32GB of RAM.

\section{Typical Order Flow}
\label{sec:appendix-order-flow}

\begin{figure*}[t]
\includegraphics[width=\linewidth]{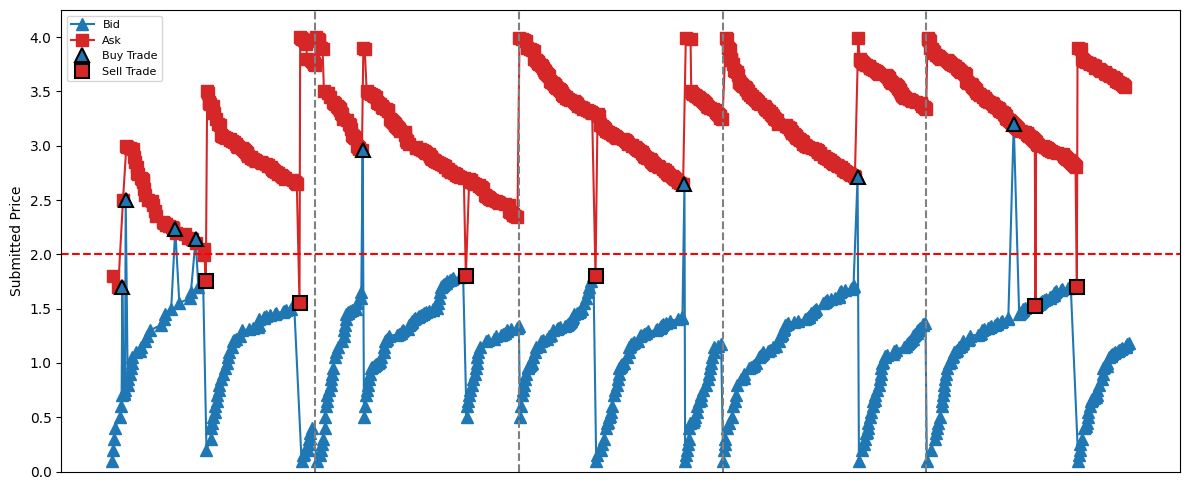}\\[1ex]
\includegraphics[width=\linewidth]{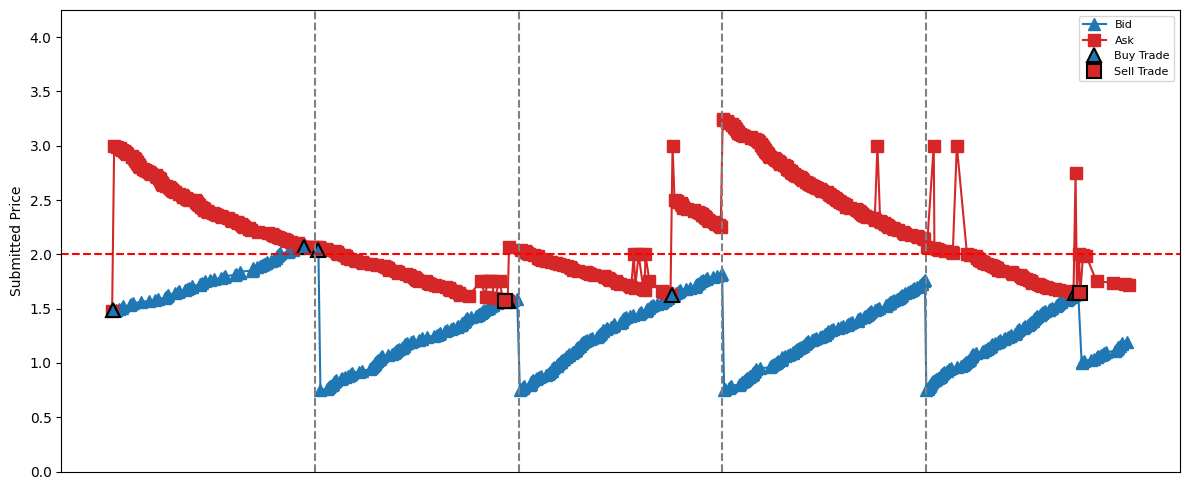}\\[1ex]
\includegraphics[width=\linewidth]{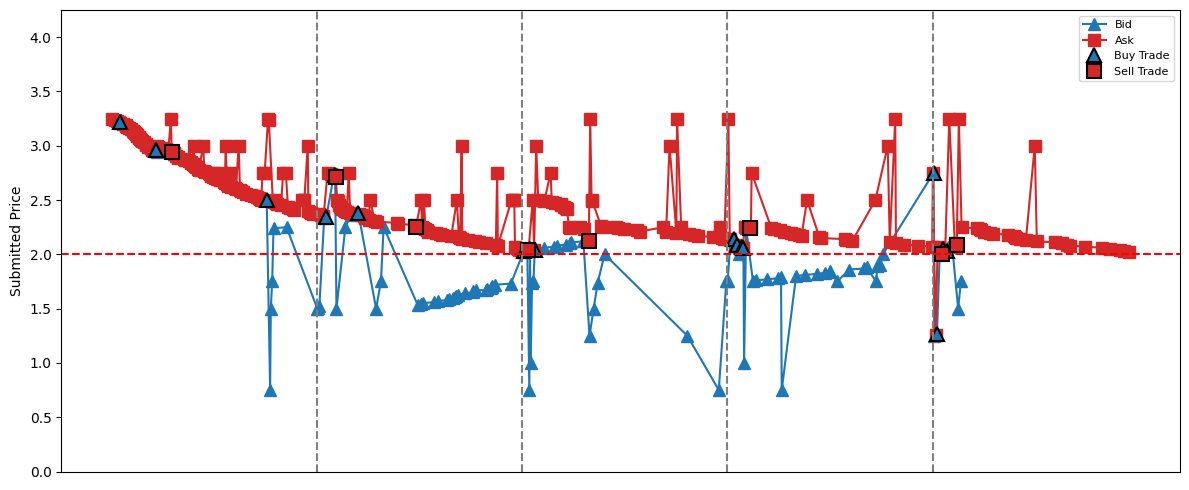}
\caption{\textbf{Individual orders submitted throughout a single sample experiment.}. Vertical dashed lines separate the 5 rounds of the experiment. The horizontal red dashed line indicates the competitive equilibrium price (\$2.00). First panel: Gemini Large; second panel: GPT Large; third panel: GPT small. Gemini Large crosses the spread once profitable trading opportunities arise, while GPT large waits until the spread goes to zero. GPT small is a mixture of these two patterns.}
\label{fig:order-flow}
\end{figure*}

Figure \ref{fig:order-flow} shows the price of each order submitted during a sample simulation for all three of our experiments. It also highlights which of these orders immediately resulted in a transaction. The picture reveals the heterogeneity in the willingness to cross the spread between models, especially between GPT Large and Gemini Large, which we refer to in the paper. Gemini Large crosses the spread once profitable trading opportunities arise, while GPT large waits until the spread goes to zero. GPT small is a mixture of these two patterns.

In the case of Gemini Large, the spread-cross moves are quite clearly distinguishable from the incremental order improvements -- a distinction we make in Section 5 of the main text.

\end{document}